\pdfoutput=1
\documentclass[12pt]{article}

\usepackage[preview]{arxiv-iaw}

\usepackage[backend=biber,style=authoryear,sorting=nyt,maxcitenames=2,mincitenames=1,maxbibnames=8,minbibnames=1,uniquelist=false,uniquename=false,giveninits=true,language=USenglish]{biblatex}
\DeclareDelimFormat{nameyeardelim}{\addspace}
\usepackage{arxiv-cyclones-macros}
\usepackage{tikz}   
\usetikzlibrary{positioning,fit,backgrounds,arrows.meta,decorations.pathreplacing}   

\makeatletter
\renewcommand{\p@subsection}{\thesection.}
\renewcommand{\p@subsubsection}{\thesection.}
\makeatother

\DeclareSourcemap{\maps[datatype=bibtex]{\map{\step[fieldsource=entrykey, match=\regexp{^knapp2025adthursat(?!.)}, final]\step[fieldsource=note, fieldtarget=addendum]}}}

\begin{document}

\title{\vspace*{-2em}Basin-Specific Changes and Intensity\\of Tropical Cyclones}

\author{Ivo Welch\\UCLA\thanks{Email: \texttt{ivo.welch@ucla.edu}. ORCID: \texttt{0000-0002-4347-7250}.  All remaining errors are my responsibility alone.}}

\date{October 2026}

\maketitle

\begin{abstract}\normalsize
  The 1990--2024 North Atlantic trend in the proportion of major-cyclone Category~3--5 fixes among all Category~1--5 fixes (the Kossin ratio, \KR) was about five times the combined Pacific trend. This contrast motivates examining their basin-specific links: aerosol loading to cyclone potential intensity, and potential intensity to observed cyclone activity. Annual increases in aerosol loading were generally associated with decreasing potential intensity both in the North Atlantic and in all three Pacific basins.  Potential intensity changes were consistently positively associated with cyclone-fix count changes in the North Atlantic, in some specifications also in the eastern North Pacific and South Pacific.  However, changes in potential intensity in the western North Pacific were significantly negatively correlated with fix-count changes.  The negative association disappeared after controlling for El Ni\~no--Southern Oscillation changes.  Furthermore, annual potential intensity changes were never significantly positively associated with \KR\ changes.
\end{abstract}

\clearpage

\noindent\textbf{Significance Statement}\medskip

\noindent Potential intensity describes how strong a tropical cyclone could become, but actual cyclone activity also depends on other conditions. This study examines two links across four ocean basins: from aerosol particles, which reduce sunlight reaching the ocean, to potential intensity; and from potential intensity to observations of strong cyclones. Annual increases in aerosol loading were generally associated with lower potential intensity. However, increases in potential intensity were not necessarily associated with increases in cyclone fixes in the Pacific.  The associations could change when El Ni\~no was taken into account. Only in the North Atlantic were both links consistently supported across the four regression specifications. These results suggest that potential intensity alone may not consistently explain Pacific cyclone activity.


\clearpage

\section{Introduction}\label{sec:intro}

\textcite{kossin2020global} measured the intensity of tropical cyclones as the proportion of major (C3+) cyclone satellite fixes among all C1+ cyclone satellite fixes (the ``Kossin ratio,'' \KR), and reported a global increase over 1979--2017 (see also the correction by \textcite{kossin2020correction}).  \textcite{jewson2020statistical} showed that this came mainly from fewer category-1 fixes outside the North Atlantic, and only to a small extent from more category 3--5 fixes.  The number of category 3--4 fixes rose significantly only in the North Atlantic.  Category 3--5 fixes declined in the three Pacific basins.  Without the North Atlantic, the global \KR\ trend was no longer significant at the 5\% level.

This paper examines how the North Atlantic differed from the other basins over 1990--2024.  The North Atlantic was the only basin in which both the number of major-cyclone fixes and the \KR\ rose significantly, although the latter was significant at the 6.3\% level.  The three Pacific basins, which carried most of the world's tropical cyclones, had smaller \KR\ trends, none significant and one (eastern North Pacific) negative.  The North Atlantic \KR\ trend was about five times that of the three combined Pacific basins.

The paper examines a simplified two-stage model within each basin.  The first stage runs from total (not only anthropogenic) aerosol loading over the cyclone genesis region to potential intensity, the thermodynamic ceiling on storm strength \parencite{bisterEmanuel1998}.  This channel has been proposed for anthropogenic aerosols in the North Atlantic \parencite{mann2006atlantic,booth2012aerosols,dunstone2013role}.  The second stage links potential intensity to observed cyclone activity \parencite{emanuel2005increasing,camargo2007use,vecchi2008whither}, measured here by cyclone-fix counts.

The analysis focuses on annual changes but first examines long-term trends. These offered little support for the first stage.  The IPCC attributes the dominant warming contribution to global surface temperature to well-mixed greenhouse gases, partly offset by aerosol cooling \parencite[Summary for Policymakers, A.1.3]{ipcc2021ar6spm}.  Over 1990--2024, total aerosol loading showed no significant trend in the North Atlantic, whereas cyclone potential intensity rose.  Across basins, the trends in potential intensity followed those in sea-surface temperature, which rose in every basin, but not those in total aerosol loading.

In annual changes, without the ENSO control, the first-stage association had the predicted sign in every basin.  Years in which aerosol loading rose tended to have lower potential intensity than the year before.  The aerosol--potential-intensity association was significant at the 10\% level or better in the North Atlantic and the eastern North Pacific, and in all four basins in ranks (which reduce the influence of the 1991 Pinatubo eruption).

Again in annual changes, the second stage was more ambiguous.  In the North Atlantic, larger annual increases in potential intensity were associated with larger increases in C1+ and C3+ fixes.  Yet in the western North Pacific, the most active basin, the association was \emph{negative}.  (The eastern North Pacific had the expected positive association, and the South Pacific showed little association.)

ENSO can move both potential intensity and cyclone activity, either as a confounder or as part of the mechanism.  In the eastern and western North Pacific, the standardized ENSO coefficients were several times larger in magnitude than the corresponding potential-intensity coefficients.  With a control for annual changes in ENSO, the negative association between changes in potential intensity and cyclone-fix counts in the western North Pacific disappeared.  In the North Atlantic, adding ENSO weakened the association between potential intensity and cyclone-fix counts, but it remained positive and significant at the 1\% level in raw changes and at the 10\% level or better in ranks.

With ENSO controlled, the activity-weighted pooled Pacific PI--fix-count association was positive but small and insignificant in both raw changes and ranks.  In the raw-change sensitivity analysis, inverse residual-variance weights instead yielded significant positive associations at both count thresholds (\appref{app:covariance}).  The South Pacific associations were significantly positive at both thresholds in ranks, but insignificant in raw changes.

With ENSO controlled, the standardized potential-intensity coefficients for C1+ and C3+ fixes were about three to nine times as large in the North Atlantic as in the three Pacific basins pooled. Without the ENSO control, they were positive in the North Atlantic and negative in the pooled Pacific.

The model makes no prediction for \KR, because increases in both C1+ and C3+ counts need not raise their ratio.  In no basin were annual increases in potential intensity significantly positively associated with increases in \KR.  In the North Atlantic, increases in potential intensity were associated with increases in both C1+ and C3+ fixes, but no significant increase in their ratio.

The paper then applies the two-stage model to the North Atlantic record back to 1945, using relative SST as a proxy for potential intensity.  On the long record, the North Atlantic relative SST and the HURDAT2 \KR\ declined from 1945 to 1973, while North American sulfur-dioxide emissions rose, and both rose over 1990--2024, after emissions had fallen.  This single reversal is consistent with the aerosol channel.  However, the timing alone cannot distinguish an aerosol effect from Atlantic multidecadal variability or other changes over the same period.  In particular, the annual changes in anthropogenic emissions showed no significant association with relative SST, so the evidence for that link in this record rested on the coinciding long-run trends.

Section~\ref{sec:methods} describes the data and methods.  Section~\ref{sec:results} updates and assesses Kossin's trend findings.  Section~\ref{sec:model} introduces the simplified model and examines the two stages of the chain.  Section~\ref{sec:na1945} examines the long North Atlantic record.  Section~\ref{sect:conclusion} concludes.  Table~\ref{tbl:summary} summarizes the results.  The supplement reports additional checks.

\section{Data and Methods}\label{sec:methods}

Supplemental sections are denoted by {\S}S.

\textbf{Sample.}  The analysis covers 1990--2024 and the four basins that geostationary satellites can view well: the North Atlantic (NA), the western North Pacific (WP), the eastern North Pacific (EP), and the South Pacific (SP).\footnote{The 1990 start follows other global cyclone studies \parencite{klotzbach2015webster,klotzbach2022trends} (\appref{app:startyear}). The Indian basins are excluded for viewing-geometry reasons (\appref{app:vzalat}).}

\textbf{Cyclone data.}  Cyclone intensities are from the Advanced Dvorak Technique--Hurricane Satellite (ADT-HURSAT) dataset (NCEI Accession 0307249, version 1.1; \textcite{knapp2025adthursat}).  It applies version 9.0 of the Advanced Dvorak Technique (ADT), an automated algorithm that estimates intensity from infrared satellite imagery, to the homogenized HURSAT-B1 version 07b geostationary imagery \parencite{knapp2007new}.  The record addresses cross-basin heterogeneity in operational best tracks and extends the record of \textcite{kossin2020global} from 2017 to 2024.  The mean \KR\ from ADT v9.0 exceeded that in the original record used by \textcite{kossin2020global}.  Most of this difference came from classifying the continuous ADT v9.0 wind estimates at the official 64- and 96-kt boundaries, whereas the original record's 5-kt grid implied effective cutoffs of 65 and 100~kt.  Despite this level difference, the annual \KR\ series from the two records were highly correlated (\appref{app:data}).

\textbf{Counting.}  The paper analyzes satellite fixes for intensity estimates.  Following \textcite{kossin2020global}, each storm's native 3-hourly track is thinned to one fix per 6-hourly synoptic time (00, 06, 12, and 18~UTC).  Each fix is classified at the Saffir--Simpson boundaries: tropical-storm strength or stronger ($\text{N}_{\text{TS}}$, $\ge 34$~kt), Category~1 or stronger (\N{1+}, $\ge 64$~kt), major (\N{3+}, $\ge 96$~kt), and, in \appref{app:reconcile}, Category~4 or stronger (\N{4+}, $\ge 113$~kt).  The Kossin ratio is $\KR \equiv \N{3+}/\N{1+}$.  Counts are basin-wide.  All fixes along a storm's track are assigned to its genesis basin in IBTrACS \parencite{knapp2010ibtracs}.  The counts are not restricted to the genesis regions defined below.  Counts are assigned to each storm's IBTrACS season and retain qualifying fixes throughout its full track, including outside the environmental averaging months below.  As a robustness check, \appref{app:storms} counts whole storms instead, each once at its lifetime-maximum intensity.  Whole-storm estimates are mostly similar but generally less statistically significant.

\textbf{Genesis-region drivers.}  Aerosol loading and the thermodynamic variables are area-averaged over each basin's main development region (MDR), its principal genesis region: NA $80$--$20^\circ$W and $10$--$20^\circ$N; EP $130$--$100^\circ$W and $10$--$20^\circ$N; WP $130$--$170^\circ$E and $5$--$20^\circ$N; SP $150^\circ$E--$170^\circ$W and $20$--$10^\circ$S.  They are then averaged over the basin's tropical-cyclone season: NA June--November, EP May--November, WP May--December, and SP November--April.  These environmental averages are matched to the corresponding storm-count season; the counts are not restricted to these months.  Aerosol loading is the total aerosol optical depth at 550~nm (\AODtau) from the MERRA-2 reanalysis \parencite{gelaro2017merra2,randles2017merra2,buchard2017merra2}.  Potential intensity (\textbf{PI}) is the Bister--Emanuel maximum potential intensity \parencite{bisterEmanuel1998}.  It is computed with the algorithm of \textcite{gilford2021pypi} from monthly MERRA-2 temperature and humidity profiles and sea-level pressure averaged over the genesis region, and from ERSST~v5 sea-surface temperature \parencite{huang2017ersst}, with the conventional reduction of the gradient wind to 80\% at the surface.  The relative sea-surface temperature (\textbf{xSST}) is the genesis-region SST minus the tropical ($30^\circ$S--$30^\circ$N) mean SST \parencite{vecchi2008whither}.  Section~\ref{sec:na1945} uses it for the historical NA analysis. ENSO is measured by the Ni\~no~3.4 index from ERSST~v5, averaged over each basin's own season, as is common in basin-level studies \parencite{camargo2005western,kossin2016past,silvers2025global}.  For the NA, Section~\ref{sec:na1945} extends the record back to 1945 with the aircraft-anchored HURDAT2 best track, CEDS anthropogenic sulfur-dioxide emissions, and volcanic stratospheric optical depth.  Table~\ref{tbl:glossary} lists the principal variables, their sources, and the exhibits that use them. Spatial maps of aerosol loading, relative SST, and PI appear in \appref{app:maps}.

\instbl{tbl:glossary}

\textbf{Aggregation.}  Worldwide count series sum the annual counts across four basins (\textbf{WW4}) or all six basins (\textbf{WW6}). Pacific sums WP, EP, and SP.  The aggregate \KR\ is the sum of C3+ fixes divided by the sum of C1+ fixes.  Equivalently, it averages basin ratios with weights proportional to each basin's C1+ count in that year.  Aggregate annual genesis-region drivers and the season-specific ENSO index instead use fixed weights: each basin's share of all 1990--2024 C1+ fixes (approximately NA 0.19, WP 0.39, EP 0.31, SP 0.10), renormalized to sum to one over the basins being combined.  Thus the ratio weights vary by year, whereas the driver weights do not. C1+ shares align the aggregate drivers with the population entering the Kossin ratio. The pooled regressions instead use TS-or-stronger shares to represent overall cyclone activity.  Trends of aggregate series are fitted after aggregation. Rows labeled ``Mean of slopes'' are the equal-weight arithmetic mean of the separately fitted basin slopes.

Pooled regressions in Tables~\ref{tbl:stages} and~\ref{tbl:stages.enso} retain separate basin-year observations and weight them by each basin's fixed share of 1990--2024 tropical-storm-or-stronger fixes: NA 0.186, WP 0.437, EP 0.242, SP 0.135.\footnote{The weights represent each basin's share of regional cyclone activity, rather than its estimated residual precision. They remain fixed across years, outcomes, and regression specifications.} Pacific uses WP, EP, and SP, renormalized to sum to one. Weighting these three regions by their geometric genesis-region box areas instead of TS-fix shares would lower the WP weight to 46.4\% (from 53.7\%) and the EP weight to 23.0\% (from 29.7\%), while raising the SP weight to 30.6\% (from 16.6\%). In the four-basin aggregation, TS weighting also gives the NA less weight than area weighting would (18.6\% versus 31.5\%). These weights give more influence to basins with more cyclone activity. Inverse residual-variance weights more than doubled the pooled Pacific raw-change Stage-2 coefficients with ENSO and made them significant (\appref{app:covariance}).

\textbf{Descriptive statistics.}  \appref{app:descriptives} reports means and variation in the levels and annual changes of these variables, distinguishing variation over time within basins from variation across basins.

\textbf{Statistics.}  Trends are Theil--Sen slopes per decade, tested with Mann--Kendall tests, as in \textcite{kossin2020global}.  Cyclone-count and ratio trend tests use the continuity-corrected normal approximation to the $S$ statistic without a tie correction to its variance.  Environmental-driver trend tests use the Kendall rank-correlation test with its tie correction.

The chain is estimated within each basin in two annual first-difference stages.  Stage~1 regresses the change in PI on the change in \AODtau.  Stage~2 regresses the change in cyclone-fix counts on the change in PI. Each stage is estimated without and with the year-to-year change in the Ni\~no~3.4 index as a control. The season average is formed before differencing. For the SP, season $t$ runs from November of year $t-1$ through April of year $t$ for ENSO, aerosol loading, and the other drivers alike.  Standard errors are Newey--West standard errors (two lags) computed on leverage-adjusted residuals: each residual is divided by one minus its leverage, as in the HC3 estimator of \textcite{mackinnon1985some}.  The adjustment accounts for the tendency of high-leverage observations, such as the 1991--92 Pinatubo changes, to have artificially small fitted residuals.  Coefficients are reported in standardized form (standard deviations of the outcome per standard deviation of the regressor), and both stages are repeated on ranks.  Stage~2 is also estimated with the change in \KR\ as the outcome.

Pooled regressions stack the basin-year observations, retaining each basin's own outcome and drivers, and add basin fixed effects.  Both the Pacific and four-basin pooled rows in the main regression tables use the fixed TS-fix shares described above. Count changes retain transitions from zero and measure additional fixes. Reported coefficients multiply fitted slopes by sd($x$)/sd($y$), using the unweighted standard deviations within the estimation sample. This reporting convention does not rescale observations before estimation. Rank regressions rank each variable within basin before applying the same activity weights. The supplemental tables also report alternative equal-observation and inverse residual-variance weights, labeled in their notes.  In pooled regressions, the Newey--West lag terms pair a basin's year only with the same basin's earlier years, and weighted regressions use weighted scores.  The standard errors assume no correlation between different basins in the same year.  The supplement provides further details and robustness checks.

\section{Kossin's Trends}\label{sec:results}

This study was inspired by \textcite{kossin2020global}, who proposed that global warming may have increased the intensity of cyclones, as measured by trends in the \KR.

\needspace{5\baselineskip}
\subsection{Half-Sample Split Differences}

Kossin showed that global six-basin \KR\ was lower in 1979--97 than in 1998--2017.  In the methods-updated data (ADT v9.0 / HURSAT-B1 v07b), this half-sample difference was no longer statistically significant over the same 1979--2017 window (\appref{app:reconcile}).  The six-basin \KR\ rise between halves was about half as large in the updated record (5 rather than 9 log points) --- mainly because the rises in the NA, SP, and South Indian basins were smaller.

Half-sample comparisons depend on the sample window and split year.  Kossin's split after 1997 coincided with a discrete jump in NA \KR.\footnote{At this paper's thresholds, mean annual NA \KR\ rose from 0.27 in 1990--97 to 0.48 in 1998--2007. Further period comparisons appear in \appref{app:reconcile}.}  For the 1990--2024 sample used here, the half-sample difference at the midpoint (after 2007) was not significant.  This comparison changes both the sample window and the split year.  Extending the record to 2024 and splitting at its midpoint (1979--2001 versus 2002--2024, excluding 1980) yielded a four-basin \KR\ increase of about 10 log points (from 0.43 to 0.47), significant at the 0.8\% level. The half-sample inference therefore depended on the window and split.

\needspace{5\baselineskip}
\subsection{Trend Estimates}\label{sec:trends}

Trend estimates avoid choosing a split year, although they still depend on the sample window.

\instbl{tbl:tc.trends}

Table~\ref{tbl:tc.trends} updates the trend statistics in \textcite{kossin2020global} and in \textcite{jewson2020statistical}.  Kossin estimated trends on three-year bins over 1979--2017.  Table~\ref{tbl:tc.trends} uses annual values over 1990--2024.

The NA had salient increases in cyclone-fix counts. The number of six-hourly major-hurricane (C3+) fixes increased from 2 in 1990 to 54 in 2024. The fitted trend implied an increase of 7.5 fixes in the annual count per decade, more than half of the fitted 1990 count (10.5 fixes; Table~\ref{tbl:tc.trends} rounds both).  Similarly, C1+ hurricane fixes increased from 38 observed fixes in 1990 to 132 in 2024, with a trend of 11 additional fixes in the annual count per decade (not statistically significant).  Finally, NA tropical-storm-or-stronger fixes increased from 283 in 1990 to 395 in 2024, with a trend of 36 additional fixes in the annual count per decade.  The Kossin ratio, the share of C3+ fixes among C1+ fixes, increased by 5.4 percentage points per decade in the NA, significant at the 6.3\% level.

Over 1990--2024, tropical-storm-or-stronger and C1+ fixes declined in all three Pacific basins, though significantly only for WP tropical-storm-or-stronger fixes.  C3+ fixes were flat in the WP and SP and fell insignificantly in the EP. All other basins had smaller \KR\ trends, never statistically significant. In the Theil--Sen estimates, the insignificant WP and SP \KR\ increases over 1990--2024 reflected declining C1 and C2 fix counts rather than increasing C3+ counts (see \textcite{jewson2020statistical}, although sample and estimator differ).  The NA slope was almost four times the four-basin worldwide slope (0.015) and five times the slope of the three Pacific basins combined (0.011). Neither WW4 nor WW6 had a significant \KR\ trend over 1990--2024. \appref{app:logslopes} separates numerator and denominator contributions using log trends, with a one-fix substitution for zero major-fix counts confined to that decomposition.

Using tropical-storm-or-stronger fixes as the denominator changed the basin comparison: the C3+ share rose most in the WP rather than the NA. The WP increase was significant, reflecting falling tropical-storm-or-stronger counts with approximately flat major-fix counts. Worldwide \KR\ inference also depended on the sample window and basin coverage.  With three-year binning, the major-share trends in both WW4 and WW6 were significant over 1979--2024 but not over 1990--2024 (\appref{app:reconcile}).  Counting whole storms instead of fixes generally weakened significance: the NA major-share trend was significant with earlier starts, but not over 1990--2024, while WW4 and WW6 remained insignificant. 

\appref{app:trendsupp} examines alternative counting units, intensity measures, trend estimators, and sample windows.  The NA increases were more persistent across these checks than the worldwide increases, although statistical significance depended on the specification and sample.  The worldwide intensity-ratio trends in Table~\ref{tbl:tc.trends} were positive, but their statistical significance depended on the denominator, counting unit, sample window, and test. The insignificant worldwide \KR\ trend over 1990--2024 was consistent in sign with Kossin's result and did not establish the absence of a trend.

\section{The Simplified Model}\label{sec:model}

Only the NA showed significant increases in both major-cyclone fix counts and \KR.  This contrast motivates examining basin-specific links from aerosols to potential intensity to cyclone activity.

\insfig{fig:dag}

Figure~\ref{fig:dag} shows a simplified model of these links.  The model considers dimming by aerosols, including anthropogenic sulfate, mineral dust, and volcanic sulfate \parencite{robock2000volcanic,booth2012aerosols,evan2016past}.  Aerosol dimming is accordingly measured by total aerosol optical depth (\AODtau), and the proposed link can be examined using both secular changes in anthropogenic emissions and the sharp 1991 Pinatubo pulse.  Total AOD combines aerosol species with potentially different radiative effects, so it is a proxy for dimming rather than a direct measure of surface solar flux. \textcite{chiacchio2011decadal} similarly analyzed total AOD in relation to European dimming and brightening. The simplification reflects that all three sources dim the surface, whereas attributing the forcing to any one of them is less secure.

The first stage links aerosol dimming to cyclone potential intensity.  Greater aerosol dimming over a basin's genesis region (its main development region, Section~\ref{sec:methods}) should reduce tropical-cyclone potential intensity over the same region \parencite{sobel2016human,rousseaurizzi2022natural}.  In CMIP5 single-forcing experiments, \textcite{sobel2019aerosol} found that aerosol cooling reduced PI by about twice as much per degree of sea-surface temperature change as greenhouse-gas warming increased it.  This motivates testing aerosol loading as a predictor of PI.

The second stage predicts that higher potential intensity raises C1+ and C3+ fix counts (see \textcite{emanuel2013influence} and \appref{app:literature}).  The simplified model makes no prediction for \KR, because increases in both C1+ and C3+ counts need not raise their ratio.  Climate-model projections assessed by \textcite{knutson2020tropical} include increases in mean cyclone intensity and the proportion of Category~4--5 storms under greenhouse warming.  Those projections concern a different forcing and different intensity metrics from the aerosol--PI--fix-count associations tested here.

Aerosol loading and potential intensity were chosen as plausible inputs to cyclone activity.  Vertical wind shear, mid-level humidity, and low-level vorticity also matter for cyclone formation, but cyclones moisten the mid-troposphere, alter the local wind profile, and are themselves vortices, so seasonal averages of these variables partly measure the storms.  These variables can mediate cyclone formation, but feedback from storms makes their effects difficult to identify from contemporaneous seasonal averages (\appref{app:shearrh} reports shear and humidity with this caveat).  Potential intensity is partly an outcome, too: storms cool the sea surface and, for about a week, warm the atmosphere above it along their tracks, which lowers the local potential intensity for one to two months \parencite{hart2007estimating}, so an active season can lower its own seasonal genesis-region PI.  Yet this feedback works against a positive association between PI and cyclone-fix counts and therefore can bias the Stage-2 estimates toward zero.

Factors that are worldwide and slow-moving, such as greenhouse warming, are not modeled separately.  They are common to all basins and change little from one year to the next.  Uniform warming raises potential intensity only weakly, because the upper troposphere warms with the surface.  The model captures the effects of these slow-moving global forcings on cyclones only through genesis-region potential intensity.

\subsection{Long-Term Trends in Variables}\label{sec:vartrends}

Table~\ref{tbl:x.trends} reports 1990--2024 trends in the genesis-region variables defined in Section~\ref{sec:methods} and Table~\ref{tbl:glossary}, alongside trends in the basin-wide \KR.

\instbl{tbl:x.trends}

\begin{description}\firmlist
\item[Aerosol Loading:]%
  Table~\ref{tbl:x.trends} shows no significant global \AODtau\ trend over 1990--2024.  Aerosol loading increases were most pronounced but not significant in the NA and EP.  Loading decreased in the WP (significant) and SP.

  Section~\ref{sec:napuzzle} discusses the discrepancy between these loading trends and the reductions in North American sulfur-dioxide emissions. MERRA-2 assimilates only total optical depth, so its model-based species split cannot establish which components accounted for the absence of an NA decline (\appref{app:merradust}, \appref{app:emis}).

\item[Temperature:]%
  The aerosol trends aligned inversely with the tropopause temperature trends, unlike the positive annual-change associations (\appref{app:stage1}).  Tropopause temperature did not rise in the NA or EP, but rose significantly in the WP.

  Consistent with global warming, the sea surface temperature trends were positive and significant in all basins.  In contrast to tropopause temperatures, the 35-year sea-surface temperature trends were unaligned with aerosol trends.

\item[Potential Intensity:]%
  Trends in potential intensity, computed using the formulation of \textcite{bisterEmanuel1998} and the implementation of \textcite{gilford2021pypi}, did not align with aerosol trends.  Instead, they aligned with sea-surface temperature trends.\footnote{At the estimated annual-change sensitivity (Section~\ref{sec:chainbreak}), aerosol trends accounted for little of the PI trends. This describes the point estimates. The uncertainty envelope reached 19\% in the WP (\appref{app:aodpitrend}).}

\item[Kossin Ratio:]%
  The final column of Table~\ref{tbl:x.trends} repeats the \KR\ slopes from Table~\ref{tbl:tc.trends}. Across the four basins, trends in the \KR\ aligned with trends in sea-surface temperature (SST) and potential intensity.
\end{description}
In trends, across the four basins, the link from aerosol loading to sea-surface temperatures (and thus potential intensity) was more tenuous than the link from potential intensity to the \KR.  (The cross-basin trend link between PI and \KR\ is not seen in annual changes, \appref{app:krpi}.)

\subsection{Trend-Adjusted Year-by-Year Changes}\label{sec:chainbreak}\label{sec:volcano}

Trending variables can be spuriously correlated. First-difference regressions test whether year-to-year changes are associated, rather than relying on shared trends in levels.

The following \emph{annual-change} tests examine associations from aerosol loading to potential intensity (``Stage~1'') and from potential intensity to cyclone-fix counts (``Stage~2'').\footnote{Long-term and short-term correlations need not share a sign. These tests concern annual associations only.}  The largest annual aerosol changes followed the 1991 Pinatubo eruption: an increase followed by a decrease.  Pinatubo was a worldwide event, so although optical depth is measured basin by basin, the raw Stage~1 annual change tests are primarily about the response of potential intensity to the worldwide Pinatubo dimming of 1991 and its clearing in 1992.\footnote{MERRA-2 combines modeled aerosols with observed optical depth. Shipborne measurements from one cruise indicated that it overestimated optical depth after Pinatubo, so the volcanic signal used here includes model error as well as observed variation \parencite{randles2017merra2}.}  Rank regressions reduce this influence.

The next two tables estimate the two stages with regressions in year-to-year changes.  Table~\ref{tbl:stages} reports them without a control for ENSO, and Table~\ref{tbl:stages.enso} with it.  Each table reports regressions on raw changes in Panel~A and on ranks in Panel~B.  Rank regressions limit the weight of outliers, especially the Pinatubo eruption.  Standardized coefficients express the association per standard deviation of $x$ in standard deviations of $y$, allowing comparisons across variables.

\needspace{9\baselineskip}
\subsubsection{Regressions without the ENSO Control}

\instbl{tbl:stages}

The stage~1 regressions show that annual changes in aerosol loading (\AOD) were negatively associated with annual changes in potential intensity (PI) in every basin, as predicted.  In raw changes, the association was significant at the 10\% level or better in the NA and EP and at the 7.9\% level in the four-basin pooled panel, but not in the WP, SP, or pooled Pacific panel.  The largest coefficient estimate appeared in the NA basin.  In ranks, the coefficient was significant in every basin and in both pooled panels.  The pooled Pacific rank coefficient was larger in magnitude than the NA coefficient.

Stage~2 regresses annual changes in cyclone-fix counts on changes in potential intensity, for C1+ and for C3+ fixes.  Larger annual increases in potential intensity were associated with larger increases in cyclone-fix counts at both thresholds in the NA and EP, as predicted.  In the WP, the association was significantly negative, contrary to the prediction.\footnote{The storm feedback on potential intensity described in Section~\ref{sec:model} can bias a positive PI--count association toward zero.  In monthly data, NA storm activity was followed by no detectable change in the region's SST or xSST.  In the WP it was followed by measurable cooling, which may contribute to that basin's negative association.  See \appref{app:shearrh}.}  In the SP and the pooled Pacific panel, it was near zero.  In ranks, the NA association remained significantly positive and the WP association significantly negative.  The EP association remained positive and significant at the 10\% level or better for both outcomes.

In sum, without the ENSO control, Stage~1 had the predicted sign in every basin, but Stage~2 showed diverging signs.  The NA had the conjectured strong positive association between potential intensity and observed cyclones, whereas the Pacific and especially the WP did not.

\needspace{5\baselineskip}
\subsubsection{The Contemporaneous Role of ENSO}

ENSO changes from one season to the next could move both potential intensity and cyclone-fix counts.  Table~\ref{tbl:stages.enso} therefore adds the year-to-year change in the Ni\~no~3.4 index as a control.\footnote{In raw-change regressions, the NA Stage-2 PI coefficients remained positive and significant at both count thresholds under every ENSO averaging window tested. The WP coefficients became significantly negative under the two preceding-winter indices (\appref{app:ensocommon}).}  ENSO temperature variation could itself be endogenous, so ENSO may be part of the mechanism rather than a confounder.  Both controlled and uncontrolled regressions are therefore informative.

These controls have physical motivations: ENSO moves where WP and SP storms form and how long they last \parencite{wang2002how,camargo2005western,vincent2011interannual},\footnote{These WP shifts also appeared in the 1990--2024 sample (\appref{app:ensopacific}).} and in the EP it acts through subsurface ocean heat that surface-based measures such as potential intensity do not capture \parencite{jin2014eastern}.

\instbl{tbl:stages.enso}

The Stage~1 regressions show that, in raw changes, the standardized ENSO coefficient was larger in magnitude than the aerosol coefficient in every basin. The ENSO coefficient was negative in the NA, WP, and SP and positive in the EP. It was negative but insignificant in both pooled panels. In ranks, it remained insignificant in the pooled Pacific panel but was significantly negative in WW4.

The model predicted a negative association between aerosol loading and potential intensity, whether or not ENSO was controlled for.  In 1991, \AODtau\ rose by more than four standard deviations of its year-to-year change in the NA, WP, and EP, and fell by more than three in 1992.\footnote{In the SP, whose season straddles the calendar year, the aerosol rise fell into the 1992 season. ENSO rose and fell in the same seasons as aerosol loading in each basin, motivating the control without establishing whether the eruption influenced ENSO.}  In unranked regressions, the pulse largely determined the slope.  The NA and EP slopes remained negative and significant, the WP and SP slopes turned positive, and the overall Pacific and WW4 slopes were negative and insignificant.  In ranked regressions, all slopes were negative, and the Pacific and WW4 slopes were significant.

In sum, the year-to-year changes were consistent with the negative aerosol influence on potential intensity found in the model experiments of \textcite{sobel2016human,sobel2019aerosol} and discussed by \textcite{wehner2024growing}.\footnote{Replacing aerosol optical depth with aerosol-only surface dimming preserved the negative NA and EP associations. The broader clear-sky surface flux, which also reflects absorption by water vapor, retained a significant association only in the NA among the individual basins. It did not preserve the EP result (\appref{app:forcing}).} The NA and EP associations were significantly negative with and without ENSO, in raw changes and ranks.\footnote{Adding one- and two-year lags of $\Delta$\AODtau\ left the estimated NA contemporaneous association nearly unchanged, but precision declined. Neither the individual lag coefficients nor the cumulative NA association was statistically significant. With two lags, the contemporaneous NA coefficient had $p=0.20$, whereas the EP coefficient remained significant. The NA sum of the current and one lagged aerosol coefficient in the PI regression was also insignificant ($p=0.134$), as was the sum with two lags ($p=0.477$; \appref{app:lags}).} With ENSO controlled, the WP and SP rank coefficients were negative but insignificant, whereas the pooled Pacific and WW4 rank coefficients remained significantly negative.

The Stage~2 regressions show that the ENSO variable was always significantly positively associated with cyclone-fix counts in the Pacific but negatively with cyclone-fix counts in the NA.  In contrast, the association between cyclone potential intensity and cyclone fixes was only significantly positive in the NA in all specifications.  (In ranks with ENSO controlled, the SP association was also significantly positive at both count thresholds.)  The coefficient estimates were about 3 to 9 times as large in the NA as in the pooled Pacific.  These results caution against using observed Pacific cyclone-fix counts as a direct proxy for potential intensity when evaluating model predictions.

These annual associations do not identify a causal aerosol effect. In the NA, potential intensity cannot be distinguished from Atlantic multidecadal variability (AMV).  The Trenberth--Shea AMV index \parencite{trenberth2006atlantic}, NA SST relative to the near-global mean, is a relative SST measured over a larger region.  It correlated 0.85 with NA PI, in levels and in annual changes.  Two otherwise identical regressions add different AMV controls: the index \emph{level} (this year's value) or its year-to-year \emph{change}. Both retain annual changes in cyclone-fix counts as the outcome and annual changes in PI and ENSO as regressors. Adding the AMV \emph{level} left the PI coefficients nearly unchanged. Adding the AMV \emph{change} instead reduced the PI coefficients for both C1+ and C3+ counts and made them statistically insignificant (\appref{app:amv}).  This collinearity is expected.  \textcite{booth2012aerosols} attribute much of twentieth-century NA multidecadal variability to aerosol forcing, so controlling for AMV changes could remove part of the mechanism under study. This multidecadal interpretation does not establish an aerosol--AMV link at annual frequency.  The annual data therefore cannot separate the potential-intensity channel from AMV.  Annual changes in aerosol loading were significantly associated with changes in NA PI, but not with changes in the AMV index.

\appref{app:krpi} shows that there was no specification (basin, unranked or ranked changes, with or without ENSO) in which a year-to-year increase in potential intensity was associated significantly with an \emph{increase} in the share of major-cyclone fixes \KR.  The only significant potential-intensity coefficient was negative (WP, without the ENSO control).  However, \KR\ moved with ENSO in the EP and in the pooled Pacific and WW4 specifications (and in the WP in ranks).

\textcite{kossin2020global,jewson2020statistical} documented the NA's distinct count and intensity-ratio trends. \textcite{emanuel2013influence} found a significant association between potential intensity and storm power only in the NA among the three basins examined, using smoothed seasonal levels. The present analysis instead tests both links from aerosol loading to potential intensity to cyclone-fix counts \emph{in annual changes} within each basin. Only in the NA were both links significant with the predicted signs in raw changes and ranks, with and without the ENSO control.

\appref{app:tercile} reports nonparametric tests that sort years by aerosol loading or potential intensity within consecutive three-year groups, reducing the influence of long-run trends. High-aerosol years had lower potential intensity in all four basins, significantly so only in the WP and SP, while high-PI years had significantly more C1+ and C3+ fixes only in the NA.

\section{The North Atlantic Basin}\label{sec:na1945}

The NA's aircraft-supported cyclone record permits extending the analysis backwards.  The NOAA National Hurricane Center's \href{https://www.nhc.noaa.gov/data/hurdat/}{HURDAT2} best track includes aircraft-reconnaissance observations from 1944 onward in a record beginning in 1851.  \appref{app:hurdat} describes the correspondence between HURDAT2 and ADT-HURSAT.  However, before satellites, HURDAT2 relied on aircraft and ship reports, so it could miss short-lived storms far from land \parencite{landsea2010impact,vecchi2011estimating}.

The 1945--2024 record uses relative SST (xSST) as a proxy for potential intensity, which is unavailable before 1980. Relative SST subtracts the tropical mean from genesis-region SST (Section~\ref{sec:methods}), so spatially uniform warming cancels. Its relation to PI is discussed by \textcite{vecchi2007remote,vecchi2008whither,ramsay2011effects}. In the modern record, annual changes in xSST and PI were correlated 0.82--0.96 across the four basins (\appref{app:xsst}).\footnote{\appref{app:carnot} reports another temperature-based alternative, the sea-surface-minus-tropopause temperature contrast (the Carnot contrast).}

The \href{https://doi.org/10.5281/zenodo.4509372}{Community Emissions Data System (CEDS)} record, available through 2019, shows that anthropogenic sulfur-dioxide emissions from North America rose until 1973 and fell thereafter.  This reversal provides a test that the 1990--2024 sample could not.  If sulfate aerosols depressed the thermodynamic potential of the genesis region, then xSST and major-hurricane fixes should have declined while emissions rose and recovered after emissions peaked.

\instbl{tbl:na1945}

Table~\ref{tbl:na1945} reports the evidence. The record is split at the 1973 emissions peak and at 1990, the start of the ADT-HURSAT sample of Section~\ref{sec:results}.

Panel~A shows the emission trends.  North American emissions rose steadily from 1945 to their 1973 peak and then fell, first slowly and then more rapidly.  European emissions peaked later, around 1979, so the European 1974--89 column mixes the end of the rise with the start of the decline.  By 2000, North American emissions had fallen about $46\%$ below their 1973 peak (European ones about $74\%$ below theirs).  By 2019, the end of the CEDS record, North American emissions were about $86\%$ below peak (European ones about $95\%$).  The decline can no longer continue indefinitely at the same absolute rate.  Emissions are also not the aerosol loading over the ocean: over 1990--2024, MERRA-2 genesis-region optical depth did not follow the emission decline (Section~\ref{sec:vartrends}).

Panel~B reports the temperature trends. In 1945--73, genesis-region SST had no significant trend, while relative SST fell (significant at the 8.1\% level). In 1974--89, SST rose and relative SST rose insignificantly. In 1990--2024, both rose significantly. Over the full record, SST rose, but the early relative-SST decline and later rise offset each other, leaving no xSST trend.

Panel~C shows that hurricane activity followed the same timing, most clearly in the major share.  \KR\ fell in 1945--73, was flat in 1974--89, and rose in 1990--2024.  \C{3+} fix counts moved the same way, but were significant only at the 10\% level.  \C{1+} fix counts showed no significant trend in any era.  The post-1990 rise in major-hurricane fixes and in \KR\ thus appeared in HURDAT2 as well as in ADT-HURSAT (Section~\ref{sec:trends}).

Panels~D and~E repeat the two stages in annual first differences. In Stage~1 (Panel~D), no coefficient was significant in 1945--73. In 1974--89, annual increases in total sulfate and volcanic optical depth were significantly associated with declines in xSST, with the sulfate record available only from 1980. In 1990--2024, the sulfate association remained significant, but the volcanic association was insignificant. Both optical-depth associations were significant over their respective full records.

Annual changes in North American anthropogenic sulfur-dioxide emissions showed no significant association with xSST changes in any era. Evidence for the anthropogenic-emissions link therefore rested on the coinciding multidecadal trends in Panels~A--C.

In Stage~2 (Panel~E), changes in xSST were significantly associated with changes in both \C{1+} and \C{3+} fix counts in 1945--73. These associations were insignificant in 1974--89 and significant again in 1990--2024. Both were significant over the full record.

The historical trend reversal does not distinguish an aerosol effect from Atlantic multidecadal variability. In the longer sample, the 1945--2024 xSST--C3+ count-change association remained significant at the 9.7\% level after controlling for AMV changes (\appref{app:amv}).  The AMV index is itself a relative NA SST, and \textcite{booth2012aerosols} attribute much of its twentieth-century variation to aerosol forcing.

In summary, the historical long-term trends were consistent with the simplified model: NA relative SST and the major-hurricane share declined while anthropogenic sulfur emissions rose, and recovered after emissions peaked. However, this single reversal could not distinguish an aerosol effect from Atlantic multidecadal variability or other changes over the same period. The annual-change results were also consistent with the simplified model over the available records: increases in sulfate loading were associated with decreases in xSST from 1980 onward, and increases in xSST were associated with more hurricane fixes over the full record beginning in 1945. Annual changes in anthropogenic emissions, however, showed no significant association with xSST.

\subsection{The North Atlantic Aerosol-Cyclone Puzzle}\label{sec:napuzzle}

IPCC AR6 \S\,11.7.1.4 \parencite{ipcc2021ar6ch11} states that the recent active tropical-cyclone seasons in the NA, the North Pacific, and Arabian basins ``cannot be explained without an anthropogenic influence,'' and that this influence ``is principally associated to aerosol forcing, with stronger contributions to the response in the North Atlantic.''\footnote{IPCC AR6 also reports the inherited SROCC assessment of \emph{medium confidence} that humans contributed to Atlantic hurricane activity since the 1970s, while noting uncertainty about the relative magnitudes of human and natural influences. See \appref{app:literature} for the supporting literature and differences in the outcomes examined.}

Panel~A of Table~\ref{tbl:na1945} documented the decline in anthropogenic sulfur-dioxide emissions underlying this proposed mechanism.  The aerosol mechanism behind this attribution is the decline of North American and European sulfur emissions \parencite{booth2012aerosols,dunstone2013role,murakami2020detected}. Over 2010--2019, the Theil--Sen emission slopes were $-80\%$ and $-89\%$ of their respective period means per decade. These use the same normalization as Table~\ref{tbl:na1945}, but a shorter window. \appref{app:emis} instead normalizes 1990--2019 slopes by fitted 1990 values.

However, MERRA-2's total aerosol optical depth over the NA genesis region did not decline in this period.  It showed no significant trend over 1990--2024 (Table~\ref{tbl:x.trends}), and after the Pinatubo years it rose, by 4\% of its mean per decade over 1993--2024 and by 8\% over 2010--2024, both significant at the 5\% level.  The decline in source-region emissions was therefore not accompanied by declining total aerosol loading over the cyclone genesis region.

Restricting attention to sulfate did not resolve this discrepancy. Between the annual values for 1990 and 2019, North American anthropogenic SO$_2$ emissions fell by $81.0\%$ and European emissions by $92.5\%$, whereas MERRA-2 sulfate optical depth over the NA genesis region fell by only $2.3\%$. These are endpoint changes over the same years, each divided by its own 1990 value, rather than the period-mean-normalized trend slopes in Table~\ref{tbl:na1945}.\footnote{Historical measurements attributed approximately half of non-sea-salt sulfate at Barbados to anthropogenic sources \parencite{savoie2002marine}, including European and North African sources. This surface estimate does not establish the source shares of genesis-region sulfate optical depth. For illustration, suppose anthropogenic sources had supplied half the local sulfate in 1990 and this component had fallen proportionally with North American emissions. Its $81\%$ decline would have lowered total sulfate by about $40.5\%$ if natural sulfate had remained unchanged. To obtain a total decline of only $2.3\%$ under those assumptions, natural sulfate would have had to increase by about $76\%$. This calculation does not establish such a natural increase: it assumes that the decline in regional emissions translated proportionally into the anthropogenic component of local sulfate loading.}

Further investigation did not reconcile the difference between the emissions and sulfate-loading changes. Over 1991--2019, annual changes in source-region sulfur-dioxide emissions did not provide a consistent cooling association with PI or significantly predict downwind AOD changes (\appref{app:emis}). Thus the annual AOD--PI evidence, which includes volcanic variation, does not by itself establish an anthropogenic-cleanup mechanism. MERRA-2 assimilates total optical depth rather than sulfate separately, and its documented anthropogenic SO$_2$ inventory holds emissions fixed after 2008 \parencite{randles2017merra2,buchard2017merra2}. Its aerosol decomposition is therefore model-based. The sulfate series could not independently establish how much anthropogenic cleanup reached the genesis region, leaving the discrepancy between declining source-region emissions and nondeclining local aerosol loading unresolved (\appref{app:merradust}).

\section{Discussion and Conclusion} \label{sect:conclusion}\label{sec:verdict}

Table~\ref{tbl:summary} summarizes the findings.  Largely documented before, the 1990--2024 trends were:
\begin{itemize}
\item%
  Cyclone fixes in the Pacific trended down, while North Atlantic fixes trended up (significantly so for tropical-storm-or-stronger and C3+ fixes).

\item%
  Total aerosol loading declined significantly only in the western North Pacific.  Total aerosol loading did not decline in the North Atlantic despite large reductions in North American and European sulfur-dioxide emissions.

\item
  Cyclone potential intensity increased significantly in the North Atlantic and in the western North Pacific.  The trend was insignificant in the other two basins.  The worldwide trend was also significantly positive.

\item
  The proportion of C3+ fixes among all C1+ fixes (\KR) increased significantly in the North Atlantic basin (at the 6.3\% level).  It increased insignificantly in the western North Pacific and South Pacific and decreased insignificantly in the eastern North Pacific.  Over its three basins, the combined Pacific \KR\ trend was insignificant.

  The statistical significance of the positive worldwide \KR\ trend was sensitive to the examined sample period.  It was insignificant over 1990--2024 but significant with an earlier start.
\end{itemize}

The annual-change tests examine the links from aerosol loading to potential intensity to cyclone activity:
\begin{itemize}
\item%
  Without ENSO control, annual aerosol-loading changes were negatively associated with PI changes in every basin, significantly so in all four basins in ranks.

  Adding the ENSO control changed the aerosol coefficients. The North Atlantic and eastern North Pacific associations remained significantly negative. The western North Pacific and South Pacific coefficients became positive and insignificant in raw changes. In ranks, all coefficients remained negative, but only the North Atlantic and eastern North Pacific basin coefficients were significant.

\item%
  Annual changes in cyclone potential intensity were positively and significantly associated with cyclone activity (both C1+ and C3+) in all four specifications only in the North Atlantic.  They were largely insignificant and sometimes contradictory in the three Pacific basins.  In the western North Pacific, the association between PI changes and cyclone-fix count changes was significantly negative without ENSO control.

\item%
  \KR\ changes were significantly positively associated only with ENSO changes, never with cyclone potential intensity changes.
\end{itemize}
The longer North Atlantic record also shows the negative association of year-to-year changes in sulfate loading with relative sea-surface temperature (a proxy for cyclone potential intensity) from 1980 on, and the positive association of relative sea-surface temperature with cyclone observations from 1945 on. Annual changes in anthropogenic sulfur-dioxide emissions, however, showed no significant association with relative sea-surface temperature.

In sum, in the annual-change investigations, the evidence supported both stages of the simplified model in the North Atlantic and (without ENSO control) in the eastern North Pacific.  For the pooled Pacific, it supported only the first stage (from aerosol loadings to potential intensity) in ranks.  The combined Pacific second stage was largely insignificant.  In the western North Pacific, the unconditional PI estimated coefficient was even unexpectedly significantly negative (tracing back to an omitted strong correlated ENSO influence).  With its larger ENSO influence, the signal-to-noise ratio in the Pacific may have made it more difficult to trace the influence of aerosols on cyclones.


\section*{Data Availability Statement}

All data are from public archives: ADT-HURSAT (NCEI Accession 0307249), HURDAT2 (NOAA NHC), ERSST.v5 and Ni\~no~3.4 (NOAA), MERRA-2 (NASA GES DISC), CEDS anthropogenic emissions, and volcanic stratospheric optical depth (GISS/CMIP7).  The replication archive for this version has reserved DOI \href{https://doi.org/10.5281/zenodo.23458629}{10.5281/zenodo.23458629}. The replication package is uploaded to the unpublished Zenodo draft and is available to editors and reviewers through the private preview link supplied in the response letter. The local replication package includes processed source data, scripts to regenerate the main paper's analysis panels, tables, and numerical estimates, and PDFs of the paper and supplemental material.

\clearpage
\section{References}

{\sloppy\printbibliography[heading=none]}   

\needspace{4\baselineskip}
\bigskip\noindent\textbf{Competing Interest Statement}\medskip

\noindent The author declares no competing interests.

\needspace{4\baselineskip}
\bigskip\noindent\textbf{Author Contributions}\medskip

\noindent All work was performed by the author.  The ADT-HURSAT dataset was provided by Kenneth~R. Knapp, Timothy~L. Olander, Christopher~S. Velden, Jennifer Gahtan, and Carl~J. Schreck.  The MERRA-2 reanalysis was produced by NASA's Global Modeling and Assimilation Office, the ERSST~v5 sea-surface temperatures by NOAA/NCEI, and the CEDS emissions inventory by the Community Emissions Data System project.  All are publicly posted.

\clearpage
\section{Exhibits}

\bigskip
\bigskip



\clearpage

\begin{glossaryfloat}

  \vspace{-2em}

  \tblcaption{tbl:glossary}{Variable Glossary}

  \footnotesize
  \setlength{\defaultaddspace}{1pt}
  \renewcommand{\arraystretch}{1.1}
  \newcommand{\glossaryentry}[2]{#1\par{\leftskip=0.6em\relax #2\par}}

  \begin{ctabular}{>{\raggedright\arraybackslash}p{3.7cm} p{7cm} >{\footnotesize}p{2.8cm}}
    \toprule
    Symbol / Reference & Description & Source / Used in \\
    \midrule
    \addlinespace
    \mc3l{\uline{\emph{Aerosol Loading}}} \\
    \glossaryentry{\AOD}{\parencite{gelaro2017merra2}} & Total genesis-region aerosol optical depth (dimensionless; the quantity MERRA-2 assimilates) --- the main-text forcing variable, entered in the Stage-1 regressions as its year-to-year change $\Delta$\AODtau\ (loading-positive: more aerosol $=$ more dimming) & MERRA-2 \newline (\texttt{= TOTEXTTAU}) \newline Tables~\ref{tbl:x.trends}, \ref{tbl:stages}, \ref{tbl:stages.enso}; \appref{app:descriptives}, \appref{app:stage1} \\
    \addlinespace
    \mc3l{\uline{\emph{Genesis-region temperatures (MDR-averaged, in \degC)}}} \\
    \glossaryentry{tropt}{\parencite{gelaro2017merra2}} & Tropopause temperature (proxy for the potential-intensity cold reservoir, the storm outflow temperature) & MERRA-2 (\texttt{= TROPT}) \newline Table~\ref{tbl:x.trends}; \appref{app:stage1} \\
    \addlinespace
    \glossaryentry{SST}{\parencite{huang2017ersst}} & Sea-surface temperature & ERSST v5 (\texttt{= sst}) \newline Table~\ref{tbl:x.trends}; \appref{app:stage1} \\
    \addlinespace
    \glossaryentry{PI}{\parencite{gilford2021pypi,bisterEmanuel1998}} & Bister--Emanuel potential intensity (primary thermodynamic measure): the thermodynamic wind ceiling, from the full MERRA-2 sounding ($V_{\text{reduc}}=0.8$) & \texttt{tcpyPI} (in kt) \newline Tables~\ref{tbl:x.trends}, \ref{tbl:stages}, \ref{tbl:stages.enso} \\
    \addlinespace
    \glossaryentry{xSST}{\parencite{vecchi2008whither,vecchi2007remote}} & Basin SST minus tropical-mean SST \newline (30$^\circ$S--30$^\circ$N); continuous PI proxy, used for the 1945--2024 record & ERSST v5\newline  (=\texttt{basin $-$ mean}) \newline Table~\ref{tbl:na1945}; \appref{app:xsst} \\
    \addlinespace
    \mc3l{\uline{\emph{Tropical-cyclone fixes (basin-wide annual counts)}}} \\
    \glossaryentry{$\text{N}_{\text{TS}}$}{\parencite{knapp2025adthursat}} & Six-hourly fixes at tropical-storm strength or stronger, $\ge 34$ kt, per year & ADT v9.0 / HURSAT-B1 v07b \newline Table~\ref{tbl:tc.trends}; pooled-regression weights \\
    \addlinespace
    \glossaryentry{\N{1+}}{\parencite{knapp2025adthursat}} & Six-hourly fixes at category-1 strength or stronger, $\ge 64$ kt, per year & ADT v9.0 / HURSAT-B1 v07b \newline Tables~\ref{tbl:tc.trends}, \ref{tbl:stages}, \ref{tbl:stages.enso} \\
    \addlinespace
    \glossaryentry{\N{3+}}{\parencite{knapp2025adthursat}} & Six-hourly fixes at category-3 (major) strength or stronger, $\ge 96$ kt, per year & ADT v9.0 / HURSAT-B1 v07b \newline Tables~\ref{tbl:tc.trends}, \ref{tbl:stages}, \ref{tbl:stages.enso} \\
    \addlinespace
    \glossaryentry{$\KR$}{\parencite{kossin2020global}} & Share of C3+ fixes among C1+ fixes & \texttt{\N{3+}/\N{1+}} \newline Tables~\ref{tbl:tc.trends}--\ref{tbl:x.trends}; \appref{app:reconcile}, \appref{app:krpi} \\
    \addlinespace
    \mc3l{\uline{\emph{Natural mode (control)}}} \\
    \glossaryentry{Ni\~no 3.4}{\parencite{huang2017ersst}} & ENSO index: equatorial-Pacific SST anomaly \newline $5^\circ$S--$5^\circ$N, $170$--$120^\circ$W & ERSST v5, Ni\~no 3.4 region control, Section~\ref{sec:chainbreak} \\
    \bottomrule
  \end{ctabular}

  \explain{Forcing and temperature variables are TC-season means over each basin's genesis region (MDR); fix counts are basin-wide. Genesis boxes and seasonal windows are specified in Section~\ref{sec:methods}. The tropical mean in xSST covers $30^\circ$S--$30^\circ$N; Ni\~no~3.4 covers $5^\circ$S--$5^\circ$N, $170$--$120^\circ$W. Stage~1 uses the annual change in total aerosol optical depth (\AODtau), a MERRA-2 reanalysis field assimilating satellite and ground-based retrievals. Sources: MERRA-2 \parencite{gelaro2017merra2}, ERSST~v5 \parencite{huang2017ersst}, and ADT-HURSAT (ADT v9.0 / HURSAT-B1 v07b) \parencite{knapp2025adthursat}.}

\end{glossaryfloat}

\clearpage

\begin{table}[ht]
  \vspace*{-2em}
  \footnotesize\addfontsize{1}
  \renewcommand{\arraystretch}{1}
  \setlength{\tabcolsep}{4pt}
  \renewcommand{\abovepanelskip}{-0.25\baselineskip}
  \renewcommand{\belowpanelskip}{-0.75\baselineskip}
  \tblcaption{tbl:tc.trends}{Per-Basin Trends, 1990--2024}

  \panel{A}{Fitted Starting Values (Theil--Sen Fitted Level in 1990)}

  \begin{ctabular}{l r r r S r r r<{\unskip\hspace{1.8em}}}
    \toprule
    & \mc3{cS}{Fixes Per Year} & \mc3c{Intensity Ratios} \\
    \cmidrule(lr){2-4} \cmidrule(lr){5-7}
    Basin & \mc1c{$\text{N}_{\text{TS}}$/yr} & \mc1c{\N{1+}/yr} & \mc1{cS}{\N{3+}/yr} & \mc1c{$\N{1+}/\text{N}_{\text{TS}}$} & \mc1c{$\N{3+}/\text{N}_{\text{TS}}$} & \mc1c{$\KR = \N{3+}/\N{1+}$} \\
    \midrule
    NA & 218 & 38 & 10 & 0.23 & 0.07 & 0.31 \\
WP & 843 & 147 & 68 & 0.17 & 0.08 & 0.49 \\
EP & 354 & 122 & 53 & 0.36 & 0.15 & 0.41 \\
SP & 196 & 34 & 14 & 0.15 & 0.07 & 0.37 \\
\quad\textdagger\,NI & 67 & 3 & 1 & 0.06 & 0.01 & 0.17 \\
\quad\textdagger\,SI & 427 & 62 & 25 & 0.13 & 0.06 & 0.47 \\
\addlinespace
Pacific & 1,394 & 283 & 125 & 0.20 & 0.09 & 0.44 \\
WW4 & 1,668 & 368 & 165 & 0.22 & 0.10 & 0.43 \\
WW6 & 2,132 & 431 & 176 & 0.19 & 0.09 & 0.43
  \\
    \bottomrule
  \end{ctabular}

  \panel{B}{Trend Slopes per Decade}

  \begin{ctabular}{l F{0} F{0} F{0} S F{3} F{3} K{3}{1.8em}}
    \toprule
    & \mc3{cS}{Fixes Per Year} & \mc3c{Intensity Ratios} \\
    \cmidrule(lr){2-4} \cmidrule(lr){5-7}
    Basin & \mc1c{$\text{N}_{\text{TS}}$/yr} & \mc1c{\N{1+}/yr} & \mc1{cS}{\N{3+}/yr} & \mc1c{$\N{1+}/\text{N}_{\text{TS}}$} & \mc1c{$\N{3+}/\text{N}_{\text{TS}}$} & \mc1c{$\KR = \N{3+}/\N{1+}$} \\
    \midrule
    NA & \tplus 36\dstar & \tplus 11 & \tplus 8\dstar & \tplus 0.001 & \tplus 0.013 & \tplus 0.054\ostar \\
WP & -117\tstar & -5 & \tplus 1 & \tplus 0.032\tstar & \tplus 0.024\tstar & \tplus 0.018 \\
EP & -17 & -15 & -9 & -0.025 & -0.017 & -0.020 \\
SP & -10 & -2 & 0 & \tplus 0.005 & \tplus 0.003 & \tplus 0.024 \\
\quad\textdagger\,NI & \tplus 5 & \tplus 2 & 0 & \tplus 0.017 & 0.000 & \tplus 0.014 \\
\quad\textdagger\,SI & -39\dstar & -2 & 0 & \tplus 0.011 & \tplus 0.006 & \tplus 0.012 \\
\addlinespace
Pacific & -152\tstar & -21 & -4 & \tplus 0.014 & \tplus 0.009 & \tplus 0.011 \\
WW4 & -133\tstar & -16 & -4 & \tplus 0.010 & \tplus 0.006\ostar & \tplus 0.015 \\
WW6 & -153\tstar & -16 & -4 & \tplus 0.012\ostar & \tplus 0.007 & \tplus 0.015
  \\
    \midrule
    WW6, 3-yr (1990--2024) & -188\ostar & -14 & -6 & \tplus 0.014 & \tplus 0.006 & \tplus 0.010 \\
WW6, 3-yr (1979--2024) & -84 & -4 & \tplus 9 & \tplus 0.010 & \tplus 0.010\tstar & \tplus 0.021\tstar \\
WW6, 3-yr (1979--2017) & -76 & -8 & \tplus 10 & \tplus 0.008 & \tplus 0.008\ostar & \tplus 0.020\dstar
  \\
    \bottomrule
    \markpvalx{(Mann--Kendall tests)}
  \end{ctabular}

  \vspace{-0.8em}

  \footnotesize

  \explain{Six-hourly fixes use TS/C1+/C3+ thresholds of 34/64/96~kt. Panel~A gives $\operatorname{median}[y_t-b(t-1990)]$, with annual-series slope $b$; ratios are fitted separately. NI's \KR\ omits five years without C1+ fixes. \textdagger\ denotes excluded viewing geometries. Pacific combines WP, EP, SP; aggregate ratios divide summed counts. Three-year WW6 rows pool six basins, annualize bin counts, and fit trends at mean observation years. The 1990--2024 row aligns three-year bins with the sample endpoint; the first bin has two years: 1990--91, 1992--94, \ldots, 2022--24 (12 bins). The 1979--2017 row follows \textcite{kossin2020global}: 1979 and 1981, 1982--84, \ldots, 2015--17 (13 bins). The extended 1979--2024 row preserves Kossin's original bins and appends 2018--20, 2021--23, and 2024 alone (16 bins). All retain the updated ADT-HURSAT data and thresholds; 1980 is excluded. Mann--Kendall tests use annual or binned observations, respectively. Full tests: \appref{app:reconcile}; whole storms: \appref{app:storms}.}

  \interpret{Among the four main basins, only the North Atlantic saw a significant increase in major-cyclone fixes, and only the North Atlantic saw a (marginally significant) rise in their share among all fixes of Category~1 or stronger.}

  {\raggedright\source{\texttt{work/1-tbl-trend4basins.R}, \texttt{work/1-tbl-triads.R} (from \texttt{work/analysis.csv}, \texttt{work/0-prepanalysis/analysis-v6-6.csv}, \texttt{work/0-prepanalysis/counts-6h.csv}, \texttt{work/0-prepanalysis/app-storm-winds.csv}, and \texttt{work/0-prepanalysis/app-storm-counts.csv}).}\par}

\end{table}

\clearpage

\begin{figure}[ht]

  \figcaption{fig:dag}{Simplified Model from Emissions to Tropical Cyclones}

  \begin{ctabular}{@{}c@{}}
    \resizebox{\textwidth}{!}{
{
\fontfamily{cmr}\selectfont
\def\normalsize{\fontsize{10}{12}\selectfont}%
\def\small{\fontsize{9}{11}\selectfont}%
\def\footnotesize{\fontsize{8}{9.5}\selectfont}%
\def\scriptsize{\fontsize{7}{8}\selectfont}%
\normalsize
\def\AODtau{AOD\kern-0.5pt{\relsize{1.5}$\tau$}}
\definecolor{cB}{gray}{0.88}       
\definecolor{cC}{gray}{0.90}       
\definecolor{cD}{gray}{0.84}       
\definecolor{cDrv}{gray}{0.72}     
\definecolor{cF}{gray}{0.94}       
\definecolor{cE}{gray}{0.80}       
\definecolor{ered}{gray}{0.35}     
\definecolor{eblue}{gray}{0}       
\definecolor{gtxt}{gray}{0.47}     
\def\ARF{ARF}
\def\ARFsu{ARF\textsubscript{SU}}
\def\ARFnsu{ARF\textsubscript{\sout{SU}}}
\def\Nca{$N_{\text{C1+}}$}
\def\Ncc{$N_{\text{C3+}}$}

\begin{tikzpicture}[
  >={Stealth[length=2.2mm]},
  font=\small,
  used/.style={draw=black, rounded corners=2pt, align=center, inner sep=4pt, minimum height=8mm, fill=#1},
  drv/.style={draw=black, line width=1pt, rounded corners=2pt, align=center, inner sep=4pt, minimum height=8mm, fill=cDrv},
  na/.style={align=center, inner sep=4pt, minimum height=8mm, text=gtxt, font=\small\itshape},
  stagelab/.style={font=\itshape\small, text=black!70, anchor=west},
  brace/.style={decorate, decoration={brace, amplitude=7pt}, draw=black!55, line width=0.8pt},
  bracelab/.style={rotate=90, anchor=center, font=\footnotesize\itshape, text=black!60},
  e/.style={->, draw=black!55, shorten >=3pt},
  ehead/.style={->, draw=eblue, line width=1.2pt, shorten >=3pt},
  eens/.style={->, draw=ered, line width=0.8pt, densely dashed, shorten >=3pt},
  edash/.style={->, draw=black!45, dashed, shorten >=3pt},
  ratio/.style={->, line width=1pt, dash pattern=on 0.2pt off 3.6pt, line cap=round, draw=black!55, shorten >=3pt},
]

\node[na] (A_volc)   at (2.5,13.65) {Volcanic\\eruptions};
\node[na] (A_nat)    at (6.2,13.65) {Other natural\\(dust, sea salt)};
\node[na] (A_anth)   at (9.9,13.65) {Anthropogenic\\emissions};

\node[na] (B_strat) at (2.5,11.35) {Stratospheric sulfate\\(volcanic)};
\node[na] (B_trop)  at (6.2,11.35) {Tropospheric sulfate\\and other aerosols};
\node[na, draw=black!45, rounded corners] (B_ghg) at (11.1,9.5) {Greenhouse gases};

\node[used=cC] (C_trf) at (7.0,8.7) {$\Delta$\AODtau\\[1pt]{\footnotesize aerosol loading}\\[1pt]{\scriptsize (total aerosol optical depth)}};

\node[drv]     (D_car)  at (3.92,6.3) {PI\\{\footnotesize potential intensity}};
\node[used=cD] (D_trop) at (10.0,6.3) {tropt\\{\footnotesize tropopause T}};
\node[used=cD] (D_sst)  at (12.2,6.3) {SST\\{\footnotesize sea surface}};

\node[used=cE, draw=black, line width=1pt] (ENSO) at (-0.8,8.7) {ENSO\\{\footnotesize (Ni\~no 3.4)}};

\node[used=cF] (F_n1) at (4.2,4.0) {\Nca};
\node[used=cF] (F_n3) at (5.8,4.0) {\Ncc};
\node[used=cF] (F_R)  at (5.0,1.7) {$\mathit{KR} = N_{\text{C3+}}/N_{\text{C1+}}$};

\begin{scope}[on background layer]
  \node[draw=black!45, dashed, rounded corners, inner sep=10pt, fit=(A_volc)(A_nat)(A_anth)] (boxA) {};
  \node[draw=black!45, dashed, rounded corners, inner sep=10pt, fit=(B_strat)(B_trop)] (boxB) {};
  \node[draw=black!45, dashed, rounded corners, inner sep=10pt, fit=(C_trf)] (boxC) {};
  \node[draw=black!45, dashed, rounded corners, inner sep=8pt,  fit=(D_car)] (boxDc) {};
  \node[draw=black!45, dashed, rounded corners, inner sep=8pt,  fit=(D_trop)(D_sst)] (boxDt) {};
  \node[draw=black!45, dashed, rounded corners, inner sep=10pt, fit=(F_n1)(F_n3)] (boxFc) {};
  \node[draw=black!45, dashed, rounded corners, inner sep=8pt,  fit=(F_R)] (boxFi) {};
\end{scope}
\node[stagelab] at (boxA.north west) {\ Exogenous drivers};
\node[stagelab] at (boxB.north west) {\ Aerosol loadings};
\node[stagelab] at (boxC.north west) {\ A. Aerosol optical depth};
\node[stagelab] at (boxDc.north west) {\ B. Potential intensity};
\node[stagelab] at (boxDt.north west) {\ Temperatures};
\node[stagelab] at (boxFc.north west) {\ C. Tropical Cyclones};
\node[stagelab] at (boxFi.north west) {\ TC intensity};

\draw[eens] (A_volc) -- (B_strat);
\draw[e] (A_anth) -- (B_trop);
\draw[e] (A_anth) -- (B_ghg);
\draw[e] (A_nat)  -- (B_trop);
\draw[e] (boxB.south) -- (C_trf);
\draw[e] (B_ghg.south) -- (boxDt.north);
\draw[e] (C_trf) -- (boxDc.north);
\draw[e] (C_trf) -- (boxDt.north);
\draw[ehead] (boxDc.south) -- (boxFc.north);
\draw[eens] (ENSO) -- (boxDc.north west);
\draw[eens] (ENSO) -- (boxDt.north west);
\draw[eens] (ENSO.south) to[out=-90,in=165] (boxFc.west);
\draw[ratio] (boxFc.south) -- (boxFi.north);
\draw[ratio] (boxDt.west) -- (boxDc.east);

\draw[brace] (-1.8,5.6) -- (-1.8,9.4);
\node[bracelab] at (-2.35,7.5) {Stage 1: aerosol $\to$ PI};
\draw[brace] (-2.7,3.5) -- (-2.7,7.0);
\node[bracelab] at (-3.25,5.25) {Stage 2: PI $\to$ cyclones};

\end{tikzpicture}
}}
  \end{ctabular}

  \explain{Shaded boxes have empirical measures. ENSO (Ni\~no~3.4) enters as a control. ENSO is shown as a separate driver for simplicity; its placement does not assume exogeneity. Factors elsewhere in the diagram may also influence ENSO, and those pathways are omitted. Unmodeled dynamic mediators are also omitted. Variables are defined in Table~\ref{tbl:glossary}.}

  \source{\texttt{plots/causal-graph-tikz.tex}; compile with \texttt{pdflatex}.}

\end{figure}

\clearpage

\begin{table}[ht]

  \tblcaption{tbl:x.trends}{Per-Basin Theil--Sen Slopes per Decade, 1990--2024}

  \begin{ctabular}{l F{3} sss F{2} F{2} sss F{2} n n ss F{3}}
    \toprule
    & \mc1{csss}{Aerosol}      & \mc2{csss}{Temperature} & \mc3{css}{Potential Intensity} & \mc1c{Cyclones} \\
    \cmidrule(lr){2-2} \cmidrule(lr){3-4} \cmidrule(lr){5-7} \cmidrule(lr){8-8}
    Basin & \mc1{csss}{\AOD}  & \mc1c{tropt}  & \mc1{csss}{SST}    & \mc1c{PI (kt)} & & \mc1{css}{} & \mc1c{$\KR$} \\
    \midrule
    NA & 0.004 & -0.07 & 0.25\tstar & 1.45\tstar & 0.35\dstar & 0.12\dstar & 0.054\ostar \\
\addlinespace
WP & -0.003\ostar & 0.31\dstar & 0.24\tstar & 1.38\dstar & -0.08 & 0.10\dstar & 0.018 \\
EP & 0.002 & 0.00 & 0.12\ostar & -0.18 & 0.12 & -0.01 & -0.020 \\
SP & -0.001 & 0.04 & 0.17\tstar & 0.77 & 0.11 & 0.03 & 0.024 \\
\addlinespace
Pacific        & 0.000 & 0.18 & 0.19\tstar & 0.84 & 0.03 & 0.05\dstar & 0.011 \\
Mean of slopes & 0.001 & 0.07 & 0.19\tstar & 0.86\tstar & 0.13\ostar & 0.06\dstar & 0.019 \\
WW4 & 0.001 & 0.11 & 0.21\tstar & 0.87\dstar & 0.11 & 0.06\tstar & 0.015
  \\
    \bottomrule
    \markpvalx{(Mann--Kendall $p$)}
  \end{ctabular}

  \explain{Drivers are genesis-region means; \KR\ uses basin-wide counts. \AOD\ is dimensionless and temperatures are in \degC. Worldwide driver trends are fitted to annual averages with fixed weights equal to each basin's share of 1990--2024 C1+ fixes; worldwide \KR\ divides summed C3+ by summed C1+ counts each year. Pacific uses WP, EP, and SP, with driver weights renormalized to sum to one. Mean-of-slopes is the equal-weight arithmetic mean of the four separately fitted basin slopes. Variable definitions are in Table~\ref{tbl:glossary}; xSST and Carnot results are in \appref{app:xsst} and \appref{app:carnot}.}

  \interpret{Aerosol loading changed little, and it did not fall over the North Atlantic despite the emission cleanup in North America and Europe.  Sea temperatures rose in every genesis region, but potential intensity rose significantly only in the North Atlantic and the western North Pacific.}

  \source{\texttt{work/1-tbl-xbasin.R} (from \texttt{work/analysis.csv}).}

\end{table}

\clearpage

\clearpage


\clearpage

\begin{table}[ht]
  \vspace{-1em}
  \tblcaption{tbl:stages}{Year-to-Year Changes without the ENSO Control: Stage~1 and Stage~2}

  \small

  \panel{A}{Raw Changes}

  \begin{ctabular}{l F{2}@{\hspace{3\tabcolsep}}>{\footnotesize}r S F{2}@{\hspace{3\tabcolsep}}>{\footnotesize}r S F{2}@{\hspace{3\tabcolsep}}>{\footnotesize}r}
    \toprule
    & \mc2{cS}{Stg 1: $\Delta$PI} & \mc2{cS}{Stg 2: $\Delta\N{1+}$} & \mc2c{Stg 2: $\Delta\N{3+}$} \\
    \cmidrule(lr){2-3} \cmidrule(lr){4-5} \cmidrule(lr){6-7}
    Specification & \mc1{c@{\hspace{3\tabcolsep}}}{$\Delta$\AOD} & \mc1{cS}{$R^2$} & \mc1{c@{\hspace{3\tabcolsep}}}{$\Delta$PI} & \mc1{cS}{$R^2$} & \mc1{c@{\hspace{3\tabcolsep}}}{$\Delta$PI} & \mc1c{$R^2$} \\
    \midrule
    \quad NA basin & -0.37\tstar & 0.13 & 0.66\tstar & 0.44 & 0.58\tstar & 0.33 \\
\addlinespace
\quad WP basin & -0.08 & 0.01 & -0.56\tstar & 0.32 & -0.58\tstar & 0.34 \\
\quad EP basin & -0.21\ostar & 0.04 & 0.34\dstar & 0.12 & 0.34\dstar & 0.12 \\
\quad SP basin & -0.25 & 0.06 & -0.12 & 0.01 & -0.02 & 0.00 \\
\addlinespace
\quad Pacific & -0.13 & 0.02 & -0.06 & 0.00 & -0.09 & 0.01 \\
\addlinespace
\quad WW4 & -0.18\ostar & 0.04 & 0.04 & 0.00 & -0.01 & 0.00
  \\
    \bottomrule
    \markpvalx{(Newey--West, lag 2, HC3-adjusted)}
  \end{ctabular}

  \panel{B}{Ranks}

  \begin{ctabular}{l F{2}@{\hspace{3\tabcolsep}}>{\footnotesize}r S F{2}@{\hspace{3\tabcolsep}}>{\footnotesize}r S F{2}@{\hspace{3\tabcolsep}}>{\footnotesize}r}
    \toprule
    & \mc2{cS}{Stg 1: $\Delta$PI} & \mc2{cS}{Stg 2: $\Delta\N{1+}$} & \mc2c{Stg 2: $\Delta\N{3+}$} \\
    \cmidrule(lr){2-3} \cmidrule(lr){4-5} \cmidrule(lr){6-7}
    Specification & \mc1{c@{\hspace{3\tabcolsep}}}{$\Delta$\AOD} & \mc1{cS}{$R^2$} & \mc1{c@{\hspace{3\tabcolsep}}}{$\Delta$PI} & \mc1{cS}{$R^2$} & \mc1{c@{\hspace{3\tabcolsep}}}{$\Delta$PI} & \mc1c{$R^2$} \\
    \midrule
    \quad NA basin & -0.35\dstar & 0.12 & 0.63\tstar & 0.39 & 0.49\tstar & 0.24 \\
\addlinespace
\quad WP basin & -0.58\tstar & 0.34 & -0.55\tstar & 0.30 & -0.56\tstar & 0.32 \\
\quad EP basin & -0.29\ostar & 0.09 & 0.39\dstar & 0.15 & 0.38\ostar & 0.14 \\
\quad SP basin & -0.47\ostar & 0.23 & -0.11 & 0.01 & 0.02 & 0.00 \\
\addlinespace
\quad Pacific & -0.48\tstar & 0.23 & -0.20 & 0.04 & -0.19 & 0.04 \\
\addlinespace
\quad WW4 & -0.45\tstar & 0.21 & -0.04 & 0.00 & -0.06 & 0.00
  \\
    \bottomrule
    \markpvalx{(Newey--West, lag 2, HC3-adjusted)}
  \end{ctabular}

  \explain{Annual changes over 1991--2024, without ENSO. Stage~2 uses six-hourly fix-count changes, including transitions from zero ($n=34$ per basin). Panel~B ranks each variable within basin. Pacific (WP, EP, SP) and WW4 include basin fixed effects and fixed 1990--2024 TS-fix-share weights in both panels. Coefficients use unweighted sd($x$)/sd($y$); pooled $R^2$ is weighted. Stars use Newey--West standard errors (lag~2, HC3 adjustment). Raw Stage~1 is largely driven by Pinatubo (Section~\ref{sec:volcano}). Table~\ref{tbl:stages.enso} adds ENSO; \appref{app:stage1} and \appref{app:stage2} report raw coefficients and alternatives.}

  \interpret{Without the ENSO control, annual increases in aerosol loading were associated with declines in potential intensity in every basin.  Annual PI changes were positively associated with changes in cyclone-fix counts in the North Atlantic, but negatively associated in the western North Pacific.}

  \source{\texttt{work/1-tbl-stages.R} (from \texttt{work/analysis.csv} and \texttt{work/0-prepanalysis/counts-6h.csv}).}
\end{table}

\clearpage

\begin{table}[ht]
  \vspace{-1em}
  \tblcaption{tbl:stages.enso}{Year-to-Year Changes with the ENSO Control: Stage~1 and Stage~2}

  \small

  \panel{A}{Raw Changes}

  {\footnotesize\begin{ctabular}{l F{2}@{\hspace{1.2\tabcolsep}}F{2}@{\hspace{3.2\tabcolsep}}>{\footnotesize}r ss F{2}@{\hspace{1.2\tabcolsep}}F{2}@{\hspace{3.2\tabcolsep}}>{\footnotesize}r ss F{2}@{\hspace{1.2\tabcolsep}}F{2}@{\hspace{3.2\tabcolsep}}>{\footnotesize}r}
    \toprule
    & \mc3{css}{Stage 1: $\Delta$PI} & \mc3{css}{Stage 2: $\Delta\N{1+}$} & \mc3c{Stage 2: $\Delta\N{3+}$} \\
    \cmidrule(lr){2-4} \cmidrule(lr){5-7} \cmidrule(lr){8-10}
    Specification & \mc1{c@{\hspace{1.2\tabcolsep}}}{$\Delta$\AOD} & \mc1{c@{\hspace{3.2\tabcolsep}}}{$\Delta$ENSO} & \mc1{css}{$R^2$} & \mc1{c@{\hspace{1.2\tabcolsep}}}{$\Delta$PI} & \mc1{c@{\hspace{3.2\tabcolsep}}}{$\Delta$ENSO} & \mc1{css}{$R^2$} & \mc1{c@{\hspace{1.2\tabcolsep}}}{$\Delta$PI} & \mc1{c@{\hspace{3.2\tabcolsep}}}{$\Delta$ENSO} & \mc1c{$R^2$} \\
    \midrule
    \quad NA basin & -0.33\tstar & -0.42\dstar & 0.31 & 0.51\tstar & -0.35\dstar & 0.53 & 0.46\tstar & -0.25\tstar & 0.38 \\
\addlinespace
\quad WP basin & 0.10 & -0.79\tstar & 0.60 & 0.16 & 0.94\tstar & 0.68 & 0.10 & 0.89\tstar & 0.67 \\
\quad EP basin & -0.27\dstar & 0.60\dstar & 0.39 & 0.03 & 0.55\tstar & 0.32 & -0.02 & 0.63\tstar & 0.39 \\
\quad SP basin & 0.24 & -0.87\tstar & 0.58 & 0.43 & 0.75\dstar & 0.27 & 0.50 & 0.70\dstar & 0.22 \\
\addlinespace
\quad Pacific & -0.10 & -0.15 & 0.05 & 0.08 & 0.75\tstar & 0.45 & 0.05 & 0.78\tstar & 0.47 \\
\addlinespace
\quad WW4 & -0.15 & -0.19 & 0.08 & 0.17 & 0.57\tstar & 0.26 & 0.13 & 0.64\tstar & 0.30
  \\
    \bottomrule
    \markpvalx{(Newey--West, lag 2, HC3-adjusted)}
  \end{ctabular}}

  \panel{B}{Ranks}

  {\footnotesize\begin{ctabular}{l F{2}@{\hspace{1.2\tabcolsep}}F{2}@{\hspace{3.2\tabcolsep}}>{\footnotesize}r ss F{2}@{\hspace{1.2\tabcolsep}}F{2}@{\hspace{3.2\tabcolsep}}>{\footnotesize}r ss F{2}@{\hspace{1.2\tabcolsep}}F{2}@{\hspace{3.2\tabcolsep}}>{\footnotesize}r}
    \toprule
    & \mc3{css}{Stage 1: $\Delta$PI} & \mc3{css}{Stage 2: $\Delta\N{1+}$} & \mc3c{Stage 2: $\Delta\N{3+}$} \\
    \cmidrule(lr){2-4} \cmidrule(lr){5-7} \cmidrule(lr){8-10}
    Specification & \mc1{c@{\hspace{1.2\tabcolsep}}}{$\Delta$\AOD} & \mc1{c@{\hspace{3.2\tabcolsep}}}{$\Delta$ENSO} & \mc1{css}{$R^2$} & \mc1{c@{\hspace{1.2\tabcolsep}}}{$\Delta$PI} & \mc1{c@{\hspace{3.2\tabcolsep}}}{$\Delta$ENSO} & \mc1{css}{$R^2$} & \mc1{c@{\hspace{1.2\tabcolsep}}}{$\Delta$PI} & \mc1{c@{\hspace{3.2\tabcolsep}}}{$\Delta$ENSO} & \mc1c{$R^2$} \\
    \midrule
    \quad NA basin & -0.30\dstar & -0.48\tstar & 0.35 & 0.46\tstar & -0.33\ostar & 0.47 & 0.32\ostar & -0.34\tstar & 0.33 \\
\addlinespace
\quad WP basin & -0.18 & -0.66\tstar & 0.61 & 0.21 & 0.99\tstar & 0.69 & 0.19 & 0.97\tstar & 0.70 \\
\quad EP basin & -0.33\dstar & 0.48\dstar & 0.32 & 0.13 & 0.58\tstar & 0.42 & 0.06 & 0.69\tstar & 0.52 \\
\quad SP basin & -0.04 & -0.63\tstar & 0.43 & 0.38\dstar & 0.76\tstar & 0.35 & 0.46\tstar & 0.68\tstar & 0.26 \\
\addlinespace
\quad Pacific & -0.38\tstar & -0.21 & 0.26 & 0.10 & 0.75\tstar & 0.52 & 0.11 & 0.76\tstar & 0.53 \\
\addlinespace
\quad WW4 & -0.35\tstar & -0.27\dstar & 0.27 & 0.18\ostar & 0.55\tstar & 0.26 & 0.17\ostar & 0.56\tstar & 0.27
  \\
    \bottomrule
    \markpvalx{(Newey--West, lag 2, HC3-adjusted)}
  \end{ctabular}}

  \explain{As Table~\ref{tbl:stages}, adding the annual change in Ni\~no~3.4 averaged over each basin's season. Panel~B ranks the outcome and both regressors within basin. Alternative seasonal indices are in \appref{app:ensocommon}.}

  \interpret{With the ENSO control, annual increases in aerosol loading were associated with declines in potential intensity in the North Atlantic and the eastern North Pacific.  Annual PI changes were positively associated with changes in cyclone-fix counts in the North Atlantic.  The activity-weighted pooled Pacific associations between potential intensity and cyclone-fix counts were insignificant in both panels.  The South Pacific associations were significant in ranks at both thresholds.}

  \source{\texttt{work/1-tbl-stages.R} (from \texttt{work/analysis.csv} and \texttt{work/0-prepanalysis/counts-6h.csv}).}
\end{table}

\clearpage

\begin{table}[ht]
  \tblcaption{tbl:na1945}{North Atlantic, 1945--2024}

  \addtolength{\tabcolsep}{2pt}
  \renewcommand{\arraystretch}{1.2}

  \begin{ctabular}{l R R R s R}
    \toprule
    & \mc1c{'45--'73} & \mc1c{'74--'89} & \mc1c{'90--'24} & \mc1c{Full} \\
    \midrule
    \mc5l{\uline{\emph{Panel A: Anthropogenic SO$_2$ emissions (\%/decade)}}} \\
    North America & 21.3\tstar & -19.3\tstar & -48.8\tstar & -12.2\tstar \\
Europe & 37.1\tstar & -19.7\tstar & -69.3\tstar & -17.0\tstar
  \\
    \addlinespace
    \midrule
    \mc5l{\uline{\emph{Panel B: Sea-surface temperature (\degC/decade)}}} \\
    MDR SST & -0.02 & 0.34\ostar & 0.25\tstar & 0.12\tstar \\
Relative SST (\,xSST\,) & -0.10\ostar & 0.06 & 0.12\dstar & 0.00
  \\
    \addlinespace
    \midrule
    \mc5l{\uline{\emph{Panel C: Hurricane activity (\%/decade)}}} \\
    \C{1+} count & -6.1 & 10.7 & 8.3 & 3.4 \\
\C{3+} count & -29.6\ostar & 0.0 & 22.0\ostar & 6.5\ostar \\
Major share (\KR) & -38.3\dstar & -1.5 & 18.0\ostar & 6.5\dstar
  \\
    \addlinespace
    \midrule
    \mc5l{\uline{\emph{Panel D: Stage 1 --- emissions and aerosol loading $\to\Delta$xSST (standardized $\beta$)}}} \\
    Anthropogenic NA SO$_2$ (CEDS) & 0.19 & -0.02 & -0.03 & 0.06 \\
Total Sulfate &  & -0.52\tstar & -0.23\tstar & -0.29\tstar \\
Volcanic AOD (GISS/CMIP7) & -0.16 & -0.46\dstar & -0.16 & -0.19\ostar
  \\
    \addlinespace
    \midrule
    \mc5l{\uline{\emph{Panel E: Stage 2 --- $\Delta$xSST $\to$ hurricane fixes (standardized $\beta$)}}} \\
    \C{1+} count & 0.40\dstar & 0.27 & 0.72\tstar & 0.55\tstar \\
\C{3+} count & 0.29\tstar & 0.22 & 0.55\tstar & 0.44\tstar
  \\
    \bottomrule
    \markpvalx{(A--C Mann--Kendall, D--E Newey--West, HC3-adjusted)}
  \end{ctabular}

  \explain{Panels~A--C use Theil--Sen slopes; percentages divide by each era's mean. CEDS SO$_2$ emissions cover North America (US, Canada, Mexico) and Western/Central Europe through 2019, so the last Panel~A era ends in 2019. Total Sulfate in Panel~D is MERRA-2 sulfate AOD over the NA genesis region, including volcanic contributions. SST is ERSST~v5 over the NA genesis region; the hurricane measures use HURDAT2 six-hourly fixes. Panels~D--E use annual first differences and standardized coefficients. MERRA-2 begins in 1980, leaving the 1945--73 sulfate AOD cell empty and restricting the 1974--89 sulfate cell to changes ending in 1981--89.}

  \interpret{While North American sulfur-dioxide emissions rose, North Atlantic relative sea-surface temperature and the HURDAT2 share of major-hurricane fixes fell.  Relative SST and the major share recovered after emissions peaked.}

  \source[2026-06-17]{\texttt{work/1-tbl-na1945.R} (from \texttt{work/analysis-NA-1945.csv}).}

\end{table}

\clearpage

\begin{table}[ht]
  \small
  \tblcaption{tbl:summary}{Summary of Findings}

  \renewcommand{\arraystretch}{1.02}

  \begin{ctabular}{l s cccc ss cc ss c}
    \toprule
    Finding                                         & NA & WP & EP & SP & Pacific & WW4 & Src                        \\
    \midrule
    \addlinespace
    \mc8l{\emph{Trends}} \\
    \quad C1+ fixes \N{1+} & \vtp & \vdot & \vtm & \vdot & \vtm & \vtm & T\ref{tbl:tc.trends} \\
    \quad Major fixes \N{3+} & \vP & \vdot & \vtm & \vdot & \vdot & \vdot & T\ref{tbl:tc.trends} \\
    \quad Major-share ratio $\KR$ & \vP & \vtp & \vtm & \vtp & \vtp & \vtp & T\ref{tbl:tc.trends} \\
    \addlinespace
    \quad Aerosol loading (total \AOD) & \vtp & \vM & \vtp & \vdot & \vdot & \vdot & T\ref{tbl:x.trends} \\
    \quad Potential intensity (PI) & \vPP & \vP & \vdot & \vtp & \vtp & \vP & T\ref{tbl:x.trends} \\
    \addlinespace
    \mc8l{\emph{Stage 1: Aerosol Loading (Dimming) and Potential Intensity}} \\
    \quad $\Delta$\AODtau\ $\to$ $\Delta$PI, no ENSO & \vMM & \vdot & \vM & \vdot & \vtm & \vM & T\ref{tbl:stages}A \\
    \quad $\Delta$\AODtau\ $\to$ $\Delta$PI, ranked & \vM & \vMM & \vM & \vM & \vMM & \vMM & T\ref{tbl:stages}B \\
    \quad $\Delta$\AODtau\ $\to$ $\Delta$PI, with ENSO & \vMM & \vtp & \vM & \vtp & \vtm & \vtm & T\ref{tbl:stages.enso}A \\
    \quad $\Delta$\AODtau\ $\to$ $\Delta$PI, ranked & \vM & \vtm & \vM & \vdot & \vMM & \vMM & T\ref{tbl:stages.enso}B \\
    \addlinespace
    \mc8l{\emph{Stage 2: Potential Intensity and Tropical Cyclones}} \\
    \quad $\Delta$PI $\to$ $\Delta$\N{1+}, no ENSO & \vPP & \vMM & \vP & \vtm & \vdot & \vdot & T\ref{tbl:stages}A \\
    \quad $\Delta$PI $\to$ $\Delta$\N{1+}, ranked & \vPP & \vMM & \vP & \vdot & \vtm & \vdot & T\ref{tbl:stages}B \\
    \quad $\Delta$PI $\to$ $\Delta$\N{1+}, with ENSO & \vPP & \vtp & \vdot & \vtp & \vtp & \vtp & T\ref{tbl:stages.enso}A \\
    \quad $\Delta$PI $\to$ $\Delta$\N{1+}, ranked & \vPP & \vtp & \vdot & \vP & \vtp & \vP & T\ref{tbl:stages.enso}B \\
    \addlinespace
    \quad $\Delta$PI $\to$ $\Delta$\N{3+}, no ENSO & \vPP & \vMM & \vP & \vdot & \vdot & \vdot & T\ref{tbl:stages}A \\
    \quad $\Delta$PI $\to$ $\Delta$\N{3+}, ranked & \vPP & \vMM & \vP & \vdot & \vtm & \vdot & T\ref{tbl:stages}B \\
    \quad $\Delta$PI $\to$ $\Delta$\N{3+}, with ENSO & \vPP & \vdot & \vdot & \vtp & \vdot & \vtp & T\ref{tbl:stages.enso}A \\
    \quad $\Delta$PI $\to$ $\Delta$\N{3+}, ranked & \vP & \vtp & \vdot & \vPP & \vtp & \vP & T\ref{tbl:stages.enso}B \\
    \addlinespace
    \quad $\Delta$PI $\to$ $\Delta\KR$ & \vdot & \vdot & \vtm & \vtp & \vdot & \vdot & \appref{app:krpi} \\
    \addlinespace
    \bottomrule
  \end{ctabular}

  \footnotesize

  \explain{Symbols encode sign and $p$-value: \vPP/\vMM\ for $p<0.01$, \vP/\vM\ for $0.01\le p<0.10$, \vtp/\vtm\ for $0.10\le p<0.5$, and \vdot\ for $p\ge0.5$. Src identifies a main-text table and panel (T) or supplement section ({\S}S). Within each Stage~1 or Stage~2 count relationship, the first two rows omit ENSO (Table~\ref{tbl:stages}, unranked then ranked), and the next two include ENSO (Table~\ref{tbl:stages.enso}, unranked then ranked). The major-share regression row is unranked with ENSO. For trends, WW4 sums four basins' counts before computing ratios and averages drivers with fixed 1990--2024 C1+ shares; Pacific uses WP, EP, and SP, renormalizing driver weights. For regression rows, WW4 and Pacific stack basin-year observations with basin fixed effects and weight observations by fixed 1990--2024 TS-or-stronger fix shares, as in Tables~\ref{tbl:stages} and~\ref{tbl:stages.enso}.}

  \interpret{Only the North Atlantic combined rising major-cyclone fixes with consistent support for both annual links across the regression specifications, although its aerosol-loading trend did not support the proposed cleanup mechanism.}
\end{table}

\clearpage

 
\end{document}